\documentclass[final]{siamltex}
\usepackage{amsmath}
\usepackage{amsfonts}
\usepackage{amssymb,amsxtra}
\usepackage{graphicx}

\def\centereps#1#2#3{\vskip#2\relax\centerline{\hbox to#1{\special
  {eps:#3 x=#1, y=#2}\hfil}}}

\title{Improved Seed Methods for Symmetric Positive Definite Linear Equations with Multiple Right-hand Sides \footnotemark[1]}

\author{Abdou M. Abdel-Rehim \footnotemark[2]
\and Ronald B. Morgan\footnotemark[3]
\and Walter Wilcox\footnotemark[4] }

\begin{document}
\bibliographystyle{plain}

\maketitle

\renewcommand{\thefootnote}{\fnsymbol{footnote}}
\footnotetext[1]{This work utilized the Baylor High-Performance Computing Cluster.  The second author was also supported by the Baylor University Sabbatical Program.}
\footnotetext[2]{Department of Physics, Baylor
University, Waco, TX 76798-7316.  ({\tt Abdou\_Abdel-Rehim@baylor.edu}).}
\footnotetext[3]{Department of Mathematics, Baylor University, Waco, TX 76798-7328 ({\tt Ronald\_Morgan@baylor.edu}).}
\footnotetext[4]{Department of Physics, Baylor
University, Waco, TX 76798-7316.  ({\tt Walter\_Wilcox@baylor.edu}).}
\renewcommand{\thefootnote}{\arabic{footnote}}

\begin{abstract}

We consider symmetric positive definite systems of linear equations with multiple right-hand sides.  The seed conjugate gradient method solves one right-hand side with the conjugate gradient method and simultaneously projects over the Krylov subspace thus developed for the other right-hand sides.  Then the next system is solved and used to seed the remaining ones.  Rounding error in the conjugate gradient method limits how much the seeding can improve convergence.  We propose three changes to the seed conjugate gradient method: only the first right-hand side is used for seeding, this system is solved past convergence, and the roundoff error is controlled with some reorthogonalization.  We will show that results are actually better with only one seeding, even in the case of related right-hand sides.  Controlling rounding error gives the potential for rapid convergence for the second and subsequent right-hand sides.

\end{abstract}

\begin{keywords} 
 linear equations, seed methods, conjugate gradient, Lanczos, QCD, multiple right-hand sides, symmetric, Hermitian
\end{keywords}

\begin{AMS}
65F10, 65F15, 15A06, 15A18
\end{AMS}

\pagestyle{myheadings}
\thispagestyle{plain}
\markboth{A. M. ABDEL-REHIM, R. B. MORGAN and W. WILCOX}{IMPROVED SEED METHODS}

\section{Introduction}

We consider a large matrix $A$ that is either real symmetric or complex Hermitian and is positive definite.  We are interested in solving the multiple right-hand side problem $Ax^j = b^j$, for $j=1, \ldots nrhs$.  Systems with multiple right-hand sides occur in many applications (see~\cite{FrMa} for some examples).  Block methods are a popular way to solve systems with multiple right-hand sides (see for example~\cite{OL80, Sa96, FrMa, bgdr, Gu07}).  However, we will concentrate on another well known approach for this problem, the seed conjugate gradient method~\cite{Sm87,SmPeMi,Jo91,SiGa,ChWa,KiMiRa,La03,Pa80B,Sa87,vdV87,ErGu}.  As usually implemented, it solves the first right-hand side with the conjugate gradient method (CG)~\cite{HeSt,Sa96} and simultaneously projects for the other right-hand side systems over the Krylov subspace that CG generates.  After the first right-hand side system is solved, CG is applied to the second right-hand side and a projection over the subspace is done for the other remaining right-hand sides.  This process continues until all the systems have been solved.  See, for example,~\cite{ChWa} for an algorithm for seed CG.  

However, seed CG does not always work as well as it should.  Seeding more than once can actually slow down the convergence.  Here we explain why this can happen and suggest seeding only once.  For problems with related right-hand sides, an approach with a projection using previous solutions is given that often gives better convergence than multiple seeding.  Next it is observed that solving the first right-hand side system beyond convergence does not always make convergence better for the other right-hand sides.  This is due to rounding error which can be controlled by reorthogonalization.  For this we use the Lanczos version of the conjugate gradient method for solving the first right-hand side.  The other right-hand sides, after being seeded, are solved with CG.  Solving the first right-hand side to high accuracy can give significant improvement in the convergence for the others.  

Section 2 gives drawbacks of seed CG with multiple seeding and shows that single seeding can be more effective.  The approach for single seeding with related right-hand sides is given.  Section 3 discusses solving the seed system past convergence and controlling the roundoff error.  Tests with large matrices from quantum chromodynamics (QCD) are in Section 4.

\section{Single Seeding for Unrelated and Related Right-hand Sides}

\subsection{Seeding more than one time}

The seed CG method is effective because the Krylov subspace generated by CG for the first right-hand side system eventually develops good approximations to eigenvectors corresponding to small eigenvalues of $A$.  For tough problems, CG cannot converge until this happens~\cite{vdSvdV,PaPavdV}.  Chan and Wan~\cite{ChWa} point out that for the other right-hand sides, the projection over this Krylov subspace removes from their residual vectors the components in the directions of these eigenvectors (we call these components the small eigencomponents).  So when CG is applied to the second right-hand side system, there is no need for the Krylov subspace to develop approximations to these eigenvectors corresponding to small eigenvalues.  This is fortunate, because with a starting vector so weak in these eigencomponents, it would be a long wait.  A problem is that if the third system is seeded with a projection over this subspace that lacks the small eigenvectors, the small eigencomponents in the residual may be raised back up.  Then when this third system is solved, CG needs to generate these eigenvectors and converges slowly.  We now give an example of this.

{\it Example 1.} Let the matrix be diagonal with $n=5000$ and with diagonal entries distributed randomly on the interval [0,1] and ordered by size.  We use eight right-hand sides with elements distributed random normal(0,1).  All right-hand sides are solved to relative residual tolerance of $10^{-8}$.  The relative residual convergence curves are shown in Figure 2.1.  The first right-hand side system takes 617 iterations, but then the second requires only 269, because the seeding has reduced the smallest residual eigencomponents.  Figure 2.2 shows the absolute values of the 25 smallest eigencomponents of the residual after it has been seeded during solution of all the previous right-hand sides.  Since the first right-hand side has not been seeded, all eigencomponents (shown with asterisks) are fairly large.  The second right-hand side has its smallest eight residual eigencomponents (circles) reduced below $10^{-6}$ by the seeding.  This significantly helps convergence.  After the first seeding, the eigencomponents for the third right-hand side are similar to those shown for the second right-hand side.  However, as Figure 2.2 shows, they have gone up after the second seeding.  Most of the smallest eigencomponents (shown with diamonds) have magnitude between $10^{-4}$ and $10^{-2}$.  Solution of the third system then takes 598 iterations.   Eigencomponents for other right-hand sides, are usually not reduced enough to effectively deflate them out (right-hand side seven is an exception).  Table 2.1 has the matrix-vector products and has a count of the vector operations of length $n$ (such as dot products and daxpy's).
While multiple seeding reduces the total number of matrix-vector products compared to not seeding from 4935 to 3978, the number of vector operations more than doubles.

\begin{figure}
\includegraphics[width=4in]{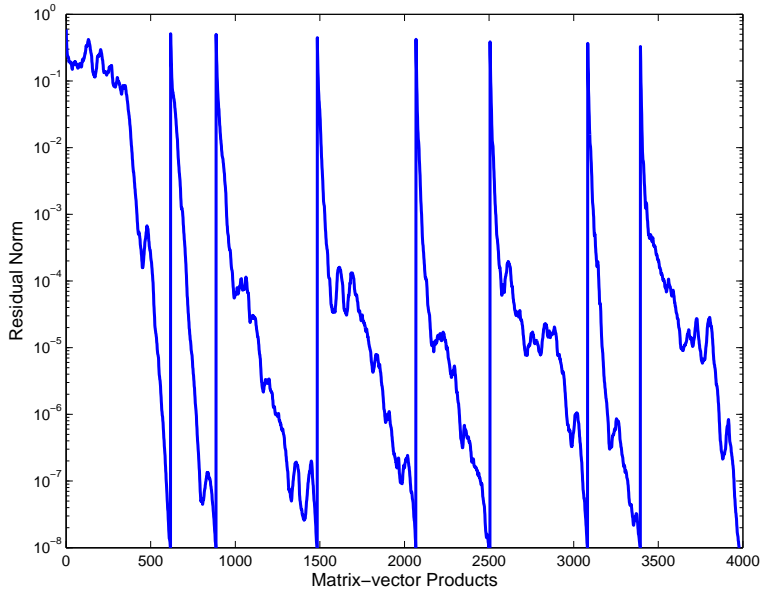}
\caption{Convergence for eight right-hand sides.  Seed with solution of every right-hand side.}
\end{figure}

\begin{figure}
\includegraphics[width=4in]{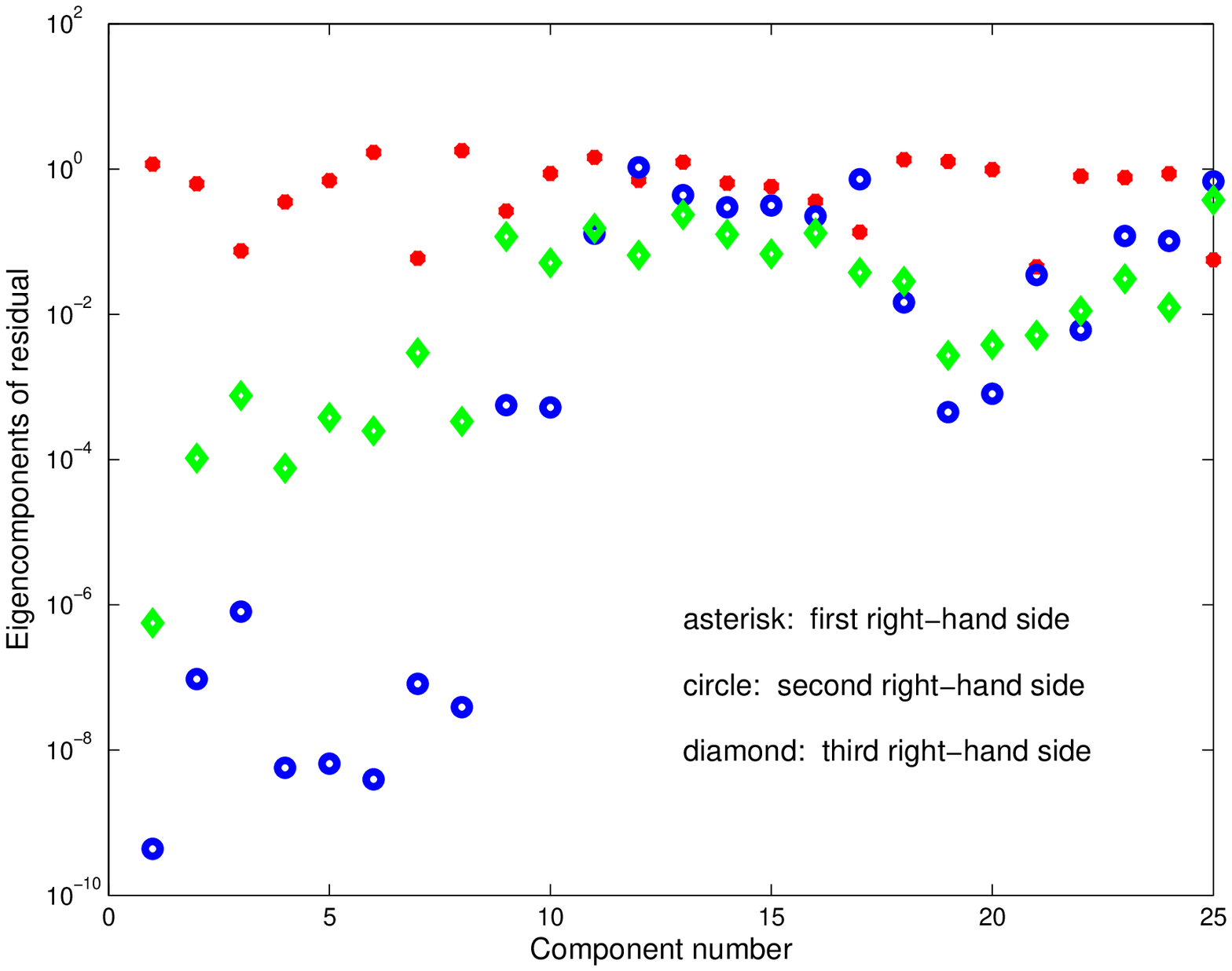}
\caption{First 25 eigencomponents of residual vectors.  The second and third right-hand sides are shown after all seeding.}
\end{figure}

\subsection{Single seeding}

We suggest seeding only one time, while solving the first right-hand side.  This can reduce the small eigencomponents for all right-hand sides, so they are prepared for the CG step.  It also significantly reduces the seeding expense.  

{\it Example 2.}  We use the same matrix and eight right-hand sides as in Example 1.  The seeding is only done one time during the solution of the first right-hand side.  The residual components after seeding of the second through eighth systems all have small eigencomponents that are similar to those of the second system in Figure 2.2.  The convergence for these systems is shown in Figure 2.3 and is much more consistent than the convergence with multiple seedings in Figure 2.1.  The total number of matrix-vector products with single seeding is 2439 compared to 3978 with multiple seeding.  Table 2.1 shows that single seeding reduces the number of matrix-vector products compared to not seeding without increasing the vector operations.

\begin{figure}
\includegraphics[width=4in]{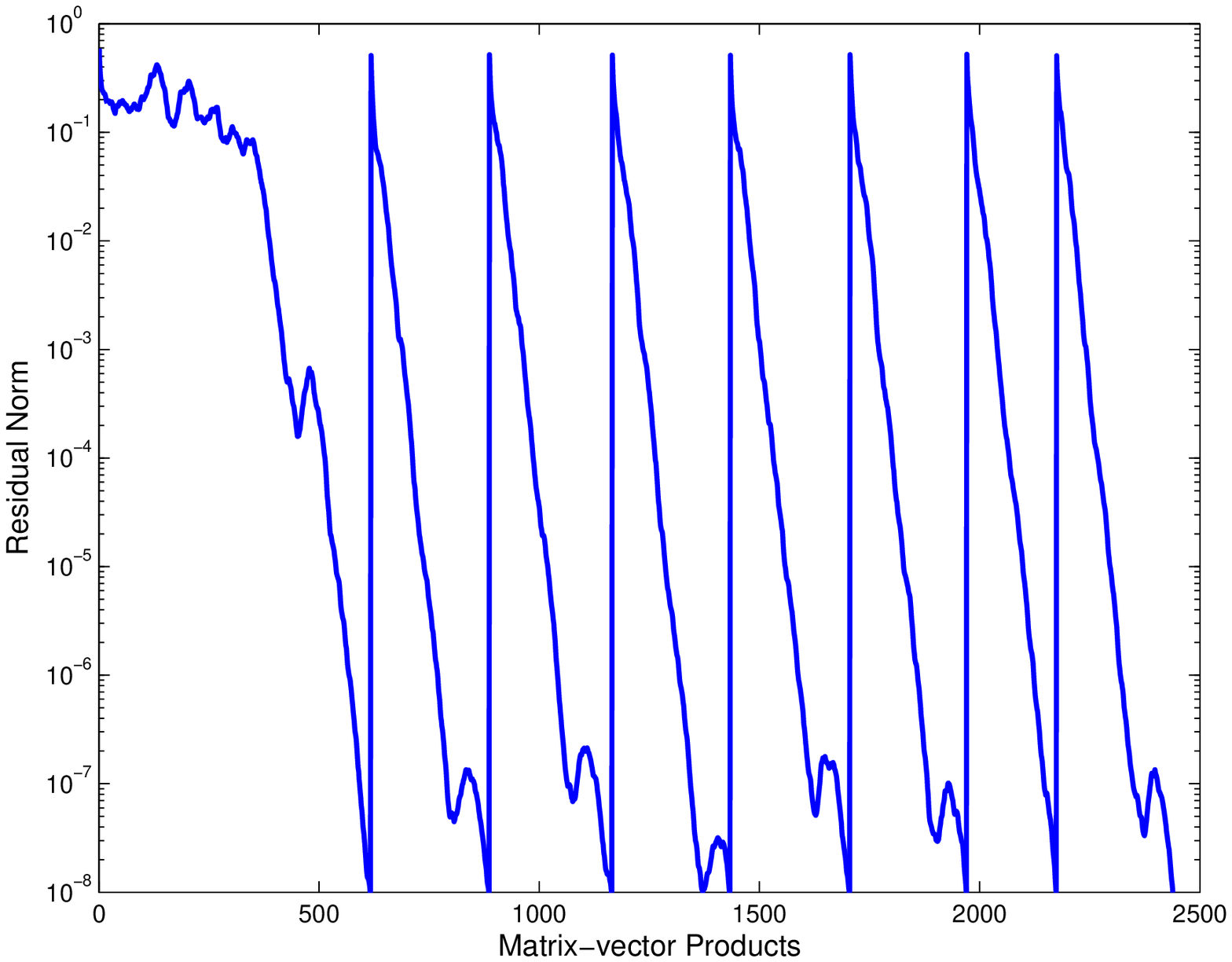}
\caption{Convergence for eight right-hand sides.  Seed only during solution of the first right-hand side.}
\end{figure}

\begin{center}
\begin{table}
\caption{Comparison of Multiple and Single Seeding }

\begin{center} \footnotesize
\begin{tabular}{|c|c|c|}  \hline\hline
                   & matrix-vector  & vector      \\
                   & products       & operations  \\ \hline
no seeding         &  4935     & 34500    \\ \hline
multiple seeding   &  3978     & 84000    \\ \hline
single seeding     &  2439     & 34300    \\ \hline
\hline

\end{tabular}
\end{center}
\end{table}

\end{center}

\subsection{Related right-hand sides}

A benefit of multiple seeding is that it can be helpful for the case of related right-hand sides.  Since systems with related right-hand sides have related solutions, projecting over a Krylov subspace containing the solution of one right-hand side reduces the residual for the next one.  Langou~\cite{La03} suggests seeding a system with only a few of the preceeding ones (a ``sliding window").  Our goal is to get the benefit of having solved related systems even when single seeding.  After solving each right-hand side except the first, we want to use that solution to help each of the remaining right-hand sides.  However, simply projecting over the solution may raise the small eigencomponents.  Instead we project over the correction to the solution found during the conjugate gradient step.  This is the difference between the solution after CG is applied and the approximate solution before CG.  This CG correction vector will generally have small components in the directions of the small eigenvectors because these eigencomponents are knocked down by the seeding before CG is applied.  Here is the algorithm for this.  

\vspace{.10in}
\begin{center}
\textbf{Single Seeding of Related Right-hand Sides}
\end{center}
\begin{itemize}
\item[1.]  \textit{Initialize.}  The right-hand sides are $b^j$ for $j=1, \ldots, nrhs$.  Here, we assume no initial guesses.  Let $x^1_{bcg} = 0$ (the subcript is for ``before cg") and $r^1_{bcg} = b^1$.  Let $j = 1$.

\item[2.]  \textit{Solve first system and seed.}  Apply CG to the first system and seed all other systems.  The solution to the first system is $x^1$.  

\item[3.] \textit{Project for the remaining right-hand sides using the solution vector.}  Form the vector $x^j_{cg} = x^j - x^j_{bcg}$ (this is just the portion of the solution found by the CG iteration).   Do a Galerkin projection over that single vector $x^j_{cg}$ for the remaining systems.  Specifically, for $i$ from $j+1$ to $nrhs$, let the current approximate solution of the $i$th system be $x^i_{bpr}$ and the current residual be $r^i_{bpr}$.  Then find $\delta = (x^j_{cg})^* r^i_{bpr} / (x^j_{cg})^* r^j_{bcg}$ and form the new approximate solution and residual for the $i$th system $x^i_{new} = x^i_{bpr} + \delta x^j_{cg}$ and $r^i_{new} = r^i_{bpr} - \delta r^j_{bcg}$.     

\item[4.] \textit{Solve the next system.}  Let $j = j + 1$.  Then for the $j$th system, let the approximate solution and residual before CG be $x^j_{bcg}$ and $r^j_{bcg}$.  Apply CG and let the solution to the $j$th system be $x^j$.  If there are more right-hand sides, go to step 3.

\end{itemize}
\vspace{.15in}

{\it Example 3.}  We use the same matrix as in the earlier examples.  The eight right-hand sides are chosen so that they become increasingly dependent on the previous ones.  Specifically, they are chosen to be $b^j = b^{j-1} + (.2)^{j-1} * u^j$, where $b^1$ and all $u^j$ have elements distributed random normal(0,1).  Figure 2.4 has relative residual convergence curves for the eight right-hand sides using multiple seeding and using single seeding with previous solution projections.  Both methods are able to take advantage of the relation between right-hand sides and begin the CG step with increasingly small residuals.  However, single seeding gives better results.  It is also less expensive after solution of the first right-hand side, because then it projects over a single vector while multiple seeding projects over a whole Krylov subspace.  We note that multiple seeding can do better once the right-hand sides are very close to each other.  For the eighth, multiple seeding takes only 44 matrix-vector products versus 70 for single seeding.  If a ninth right-hand side is solved, it takes 22 for multiple versus 55 for single.  The multiple seeding can reduce more of the small eigencomponents than can single seeding.  And though projecting more than once in multiple seeding still can raise up the size of these components, they are small enough, because CG does not need to converge very far.  

\begin{figure}
\includegraphics[width=4in]{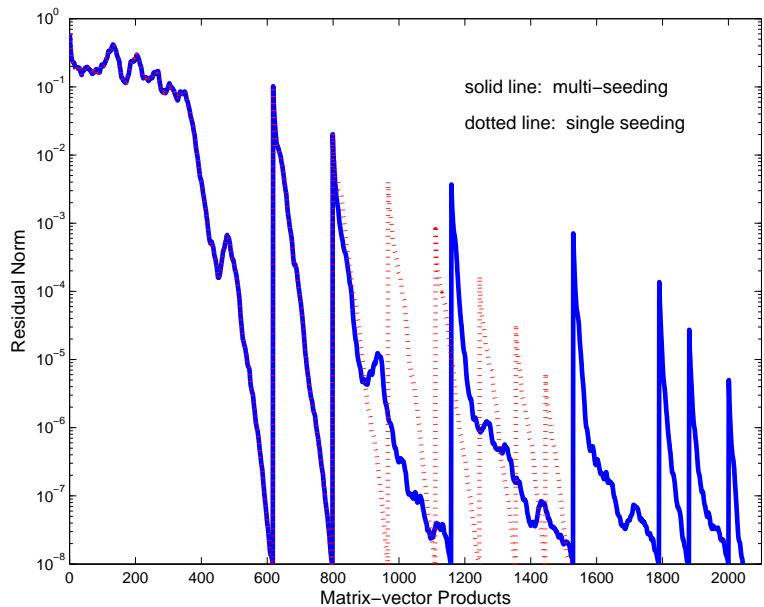}
\caption{Convergence for eight related right-hand sides.  Comparison of multi-seeding with seeding only during solution of the first right-hand side.}
\end{figure}

\section{Solving the First Right-hand Side Past Convergence}

In Examples 2 and 3, seeding significantly improves convergence because the small eigenvalues are essentially deflated out.  Recall in Figure 2.2 that eight eigencomponents for the second system are reduced to $10^{-6}$ or below.  However, we would like to reduce more components and get even faster convergence.  One way to attempt this is to solve the first right-hand side further.  The next example discusses this.

{\it Example 4.}  For the same matrix and two unrelated right-hand sides, we test solving the first right-hand side to greater accuracy.  Table 3.1 has the results for the convergence of the second right-hand side with the first solved for different numbers of iterations.  At 600 matrix-vector products for the first, the number of matrix-vector products for the second is at 270.  However, with 800 for the first, the second goes up to 471.  With 1200 for the first, the second is down to 232, but it goes back up with more for the first.  To see why this happens, we look at the magnitudes of the smallest three eigencomponents of the residual for the second right-hand side as it is being seeded.  Figure 3.1 shows that these components drop dramatically from 500 to 600 iterations.  That is when the Krylov subspace generated by CG develops good approximations to the eigenvectors corresponding to these three eigenvalues.  However, because of roundoff error, these components grow again and actually oscillate.  

\begin{table}

\caption{Solution of the second right-hand side when the first is solved past convergence}

\begin{center}
\begin{tabular}{|c|c|c|}  \hline\hline
      mvp's for 1st rhs   & 2nd rhs - 		 & 2nd rhs -     \\
				  & no reorthog. of 1st  & reothog. 1st  \\ \hline
\hline
no proj  & 617  & 617   \\ \hline
500      & 612  & 613   \\ \hline
600      & 270  & 271   \\ \hline
700      & 289  & 196   \\ \hline
800      & 471  & 164   \\ \hline
1000     & 488  & 102   \\ \hline
1200     & 232  & 64    \\ \hline
1500     & 472  & 54    
         
\\ \hline \hline

\end{tabular}
\end{center}


\end{table}

\begin{figure}
\includegraphics[width=4in]{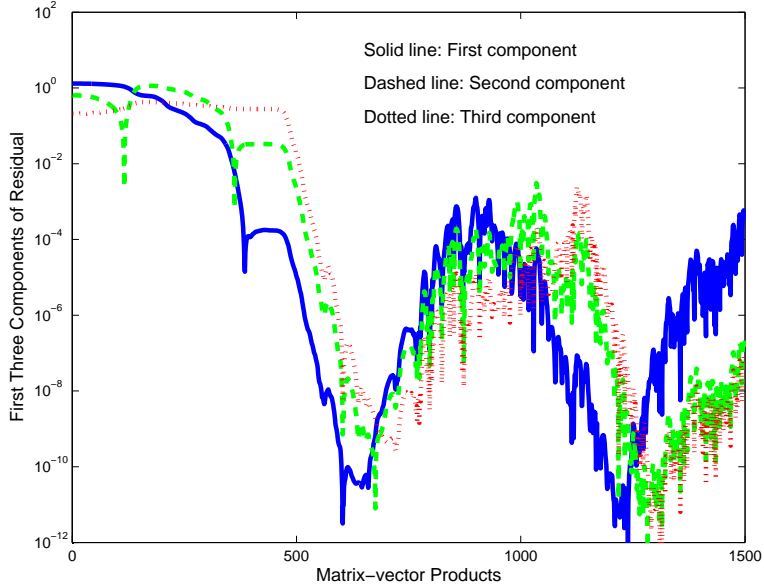}
\caption{Eigencomponents for seed CG without reorthogonalization.}
\end{figure}

Roundoff error can be controlled with reorthogonalization.  For this we use the Lanczos implementation of the conjugate gradient method.  Several ways of reorthogonalizing the Lanczos vectors have been proposed, including selective reorthogonalization~\cite{PaSc}, periodic reorthogonalization~\cite{Gr81,Stv2}, and partial reorthogonalization~\cite{Si84,Si84b,WuSi,Stv2}.  We will use periodic reorthogonalization, which is fairly easy to implement.  For this we always reorthogonalize two consecutive Lanczos vectors against all previous ones, but do this only at regular intervals during the iteration.  We call how often this is done the frequency of reorthogonalization.  See~\cite{WuSi,Stv2} for discussion of reorthogonalization and for why consecutive vectors need to be reorthogonalized.
The extra storage for saving the Lanczos vectors is only needed during solution of the seed system.  Storage can be allocated/deallocated at run time.

The following algorithm is for first right-hand side solution and seeding of the other right-hand sides.  We use LU factorization of the Lanczos tridiagonal matrix.  A QR~\cite{Sa84} or LQ factorization~\cite{PaSa} should be used for solution of indefinite systems.  

\vspace{.10in}
\begin{center}
\textbf{Lanczos Version of Single Seed CG with Periodic Reorthogonalization}
\end{center}
\begin{itemize}
\item[1.]  \textit{Start.}  Choose $m$, the maximum size of the subspace, and choose the frequency of reorthogonalization.  The number of right-hand sides is $nrhs$.  For $j=1, \ldots nrhs$, set the approximate solution denoted by $x^j$ equal to the initial guess (which may be the zero vector) and compute the initial residual $r_0^j$.  

\item[2.]  \textit{First Lanczos iteration.}  $\beta_0 = ||r_0^1||$; $v_1 = r_0^1/\beta_0$; $f = A v_1$; $\alpha_1 = v_1^* f$; $f = f - \alpha_1 v_1$; $\beta_{1} = ||f||$.

\item[3.] \textit{Linear equations for first iteration.} $\delta_1 = \alpha_1$; $w_1 = v_1 / \delta_1$; $\zeta_1 = \beta_0$; $x^1 = x^1 + \zeta_1 w_1$.

\item[4.] \textit{Other right-hand sides for first iteration.} For $j=2, \ldots nrhs$, $\eta^j = v_1^* r^j_0$; $x^j = x^j + \eta^j w_1$.  Set $i = 2$.

\item[5.]  \textit{Lanczos iteration.}  $f = A v_i - \beta_{i-1} v_{i-1}$; $\alpha_i = v_i^* f$; $f = f - \alpha_i v_i$.  

\item[6.] \textit{Linear equations.}  $\gamma_{i-1} = \beta_{i-1}/\delta_{i-1}$; $\delta_i = \alpha_i - \gamma_{i-1} \beta_{i-1}$; $w_i = (v_i - \beta_{i-1} w_{i-1} ) / \delta_i$; $\zeta_i = - \gamma_{i-1} \zeta_{i-1}$; $x^1 = x^1 + \zeta_i w_i$.

\item[7.] \textit{Other right-hand sides.} For $j=2, \ldots nrhs$, $\eta^j = v_i^* r^j_0 - \gamma_{i-1} \eta^j$; $x^j = x^j + \eta^j w_i$.

\item[8.] \textit{Reorthogonalization.} If $i$ is a multiple of the frequency of reorthogonalization, then reorthonormalize $v_i$ against $v_1, \ldots v_{i-1}$ and reothogonalize $f$ against $v_1, \ldots v_{i}$.  

\item[9.]  \textit{Finish iteration and go to next iteration.} $\beta_i = ||f||$; $v_{i+1} = f / \beta_i$.  If $i < m$, set $i = i+1$ and go to step 5.

\end{itemize}
\vspace{.15in}

{\it Example 5.}  Continuing Example 4, we now use seed CG with reorthogonalization.  The frequency of reorthogonalization is set to 50.  The results are actually the same with frequency of 75, but are much worse with 100.  The last column of Table 3.1 has the results for solving the second right-hand side.  They show that a dramatic decrease in the number of iterations can be achieved with this approach.  For instance, if the iteration for the first right-hand side is run to 1200, the second right-hand side can be solved in only 64 iterations.  This compares to 270 iterations if the first is solved with only 600 iterations of non-reorthogonalized seed CG.  Figure 3.2 has the small eigencomponents for the second right-hand side after seeding with the reorthogonalized method.  It shows that there are many small eigencomponents after running the first system for 1200 iterations.  Solving the first right-hand side past convergence may not pay off if only a few right-hand sides are solved.  However, for this example it only takes four right-hand sides for this approach to reduce the number of matrix-vector products.   

\begin{figure}
\includegraphics[width=4in]{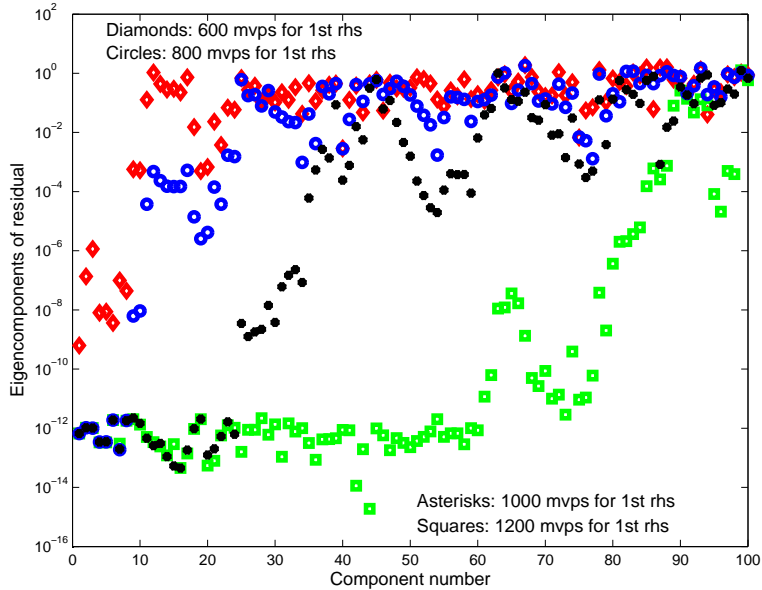}
\caption{Eigencomponents for the second system after seeding with reorthogonalization.}
\end{figure}

\section{Example from QCD}

Many problems in lattice quantum chromodynamics (QCD) have large systems of equations with multiple right-hand sides.  For example, the Wilson-Dirac formulation~\cite{Frommer,Qcdconf2} and overlap fermion~\cite{NaNe,FrOverlap2} computations both lead to such problems.  Preconditioning is usually not helpful for QCD problems.  The matrices have complex entries.  They generally are non-Hermitian, but it often is possible to change into Hermitian form.  We use the $A^*A$ approach that gives a Hermitian positive definite system (another option is the $\gamma_5 A$ approach~\cite{Frommer} which creates an indefinite matrix).  So we multiply the system by the Hermitian transpose of the matrix.  Every CG iteration requires two matrix-vector products.  

{\it Example 5.} We choose a large QCD matrix of size $n = 1.5$ million.  More specifically, we have a $20^3$x $32$ quenched Wilson configuration at $\beta=6.0$ and with $\kappa$ value  set near to $\kappa$-critical, which makes it a difficult problem.  For this application, we generally solve 150 right-hand sides per matrix, but here we just give results for two that correspond to Dirac and color indices 1,1 and 1,2 respectively.  Standard CG requires about 2500 iterations or 5000 matrix vector products, however solving the first right-hand side past this point gives better results for the second.  Figure 4.1 shows the results for the second right-hand side with regular CG and after seeding with the first solved from 5000 matrix-vector products up to 8000.  In the legend, Seed-CG(2500,100) refers to 2500 iterations for the first right-hand side with reorthogonalization of two vectors every 100 iterations.  The results are the same with frequency of reorthogonalization 200, but degrade with reorthogonalization only every 250 iterations.  

\begin{figure}
\includegraphics[width=4in]{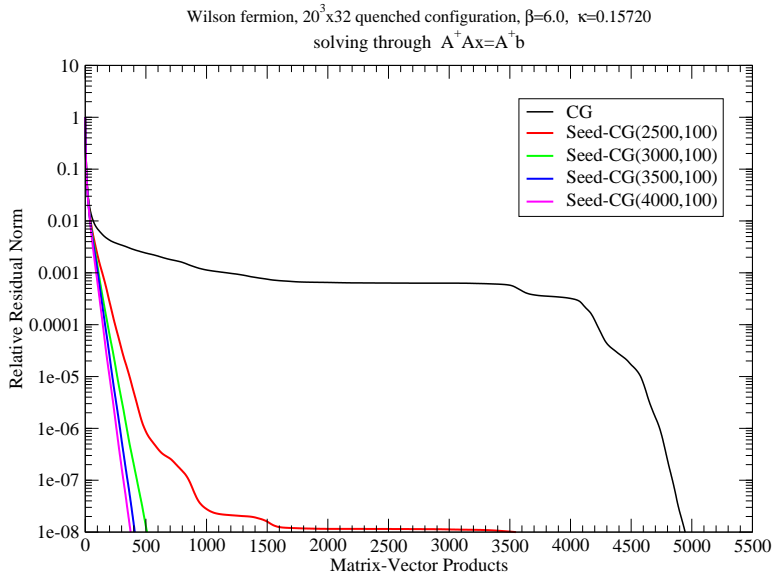}
\caption{Convergence for the second right-hand side of a large QCD matrix.  This is after being seeded by solving the first with different numbers of iterations.  Seed-CG for the first right-hand side runs from 2500 to 4000 iterations with frequency of reorthogonalization of 100.}
\end{figure}




Because of the reorthogonalization, this algorithm requires the storage of all the Krylov subspace vectors while solving the first right-hand side.  In Table 4.1, we compare the time used by CG and seed-CG for the first right-hand side. With the same number of matrix-vector products, the  cost increases from 227 seconds for CG to 320 for seed-CG with reorthogonalization every 200 iterations.  The cost is greater because of both the reorthogonalization and the seeding for the second right-hand side.  The results were obtained using the high-performance cluster at Baylor University which has 8 processors per node at 2.66GHZ and 16GB RAM per node.  The run was done on 50 processors using 5 processors per node allowing for 3.2 GB RAM per process.  It was possible to allocate up to 4000 Krylov vectors (enough for 8000 matrix-vector products) using this configuration without a need for memory swapping that leads to a slower run.  It is interesting that it was possible to store the vectors needed for an efficient method, and that the CPU cost was not too large.

\begin{center}
\begin{table}
\caption{Timings for solving the first right-hand side with different reorthogonalization choices.}

\begin{center} \footnotesize
\begin{tabular}{|c|c|}  \hline\hline
  frequency of     & time in \\
reorthogonalization     & CPU seconds   \\ \hline
10       & 1253    \\ \hline
100      & 598     \\ \hline
200      & 320     \\ \hline
250      & 305     \\ \hline
no reothog. & 227  \\ \hline
\hline

\end{tabular}
\end{center}
\end{table}

\end{center}

\section{Conclusion}
For linear systems with multiple right-hand sides, the seed conjugate gradient method can often significantly reduce the number of matrix-vector products required.  We have shown that single seeding is often better than the standard multiple seeding.  An approach was given for the case of related right-hand sides that gives better results than multiple seeding unless the right-hand sides are very closely related.

We also experimented with solving the first right-hand side past convergence and showed that this has the potential to greatly reduce the number of iterations for the other right-hand sides.  Reorthogonalization of Lanczos vectors is needed for this to be effective, so storage is a concern.   

Future work could involve exploring why roundoff error in CG has a greater negative effect for seeding than it does for solution of the first right-hand side.  Developing block versions of the seed methods discussed here would also be useful.  


\bibliography{seed.bbl}

\end{document}